\documentclass[sigplan,10pt]{acmart}

\acmConference[]{}{}{}
\acmYear{}
\acmISBN{} 
\acmDOI{} 

\setcopyright{none}

\usepackage{booktabs}   
\usepackage{subcaption} 

\usepackage[frozencache,cachedir=.]{minted}

\begin{document}

\title[Serenity: Library Based Python Code Analysis for Code Completion and Automated Machine Learning]{Serenity: Library Based Python Code Analysis for Code Completion and Automated Machine Learning}


\author{Wenting Zhao}
\affiliation{
  \department{Department of Computer Science}              
  \institution{Cornell University}            
}
\email{wzhao@cs.cornell.edu}          

\author{Ibrahim Abdelaziz}
\affiliation{
  \department{Thomas J. Watson Research Center}             
  \institution{IBM Research}           
}
\email{Ibrahim.abdelaziz1@ibm.com}         

\author{Julian Dolby}
\affiliation{
  \department{Thomas J. Watson Research Center}             
  \institution{IBM Research}           
}
\email{dolby@us.ibm.com}         

\author{Kavitha Srinivas}
\affiliation{
  \department{Thomas J. Watson Research Center}             
  \institution{IBM Research}           
}
\email{Kavitha.Srinivas@ibm.com}         

\author{Mossad Helali}
\affiliation{
  \department{Department of Computer Science}             
  \institution{Concordia University}           
}
\email{mossad.helali@concordia.ca}         

\author{Essam Mansour}
\affiliation{
  \department{Department of Computer Science}             
  \institution{Concordia University}           
}
\email{essam.mansour@concordia.ca}         

\begin{abstract}
Dynamically typed languages such as Python have become very popular\footnote{https://www.techrepublic.com/article/programming-languages-pythons-growth-is-absolutely-explosive-says-anaconda-ceo-and-not-slowing-down/}. Among other strengths, Python's dynamic nature and its straightforward linking to native code have made it the de-facto language for many research areas such as Artificial Intelligence. 
This flexibility, however, makes static analysis very  hard. While creating a sound, or a soundy, analysis for Python remains an open problem, we present in this work Serenity, a framework for static analysis of Python that turns out to be sufficient for some tasks.  The Serenity framework exploits two basic mechanisms: (a) reliance on dynamic dispatch at the core of language translation, and (b) extreme abstraction of libraries, to generate an abstraction of the code. We demonstrate the efficiency and usefulness of Serenity's analysis in two applications: code completion and automated machine learning.  In these two applications, we demonstrate that such analysis has a strong signal, and can be leveraged to establish state-of-the-art performance, comparable to neural models and dynamic analysis respectively.  
\end{abstract}



\maketitle

\section{Introduction}

 Static analysis of Python is hard, due in part to features often regarded as strengths: its dynamic nature and its straightforward linking to native code.  Python is dynamically typed, so the aid static types provide to analysis of e.g. Java is not available.  Python has a dynamic object structure; methods can be freely assigned and modified complicating resolving calls.  Even basic constructs such as method calls and object creations can be ambiguous in the basic syntax.  Beyond the language itself, many of the rich collection of Python libraries, especially the math-heavy libraries used in machine learning, are implemented in native code, which makes analysis require cross-language support.  For these reasons among others, to our knowledge, there is a lack of widely-used analysis frameworks for Python, despite the value such analysis would have, for instance, for tools.
 
 However, while creating a sound, or a soundy, analysis for Python remains an open problem, we demonstrate Serenity\footnote{With apologies to Reinhold Niebuhr, "give us courage to model what must be modeled, serenity to accept what cannot be modeled, and the insight to know the one from the other."}, a framework that turns out to be sufficient for some tasks.  Beyond a relatively direct translation of Python Abstract Syntax Tree (AST) into a Control Flow Graph (CFG), Serenity exploits two basic mechanisms:
 \begin{enumerate}
     \item Reliance on dynamic dispatch at the core of language translation.  It is not possible, always, even to tell whether a construct is an object creation or a function call, and this is just one example.  Our approach to such situations is to turn them into dynamic dispatches over types representing constituent constructs.  We detail how many subtleties of Python can be modeled in this way.
     \item Extreme abstraction of libraries.  User code often makes heavy use of APIs to create and operate upon domain objects, such as arrays in numpy, but these objects are often fairly opaque to the user code.  As such, we find it often suffices to treat libraries by just tracking the objects they create and methods called upon them.  This is not, nor is it designed to be, soundy, let alone sound.  We show however that this enables useful modeling of user code.
 \end{enumerate}
 We first discuss how Python is modeled and how the library abstraction still provides a useful analysis of user code.  We then demonstrate that this analysis is useful where we focus on two applications that depend on the outputs of such analysis:
 \begin{description}
     \item[Code Completion] is a core functionality expected in all IDEs, where the goal is to suggest methods and functions to call given prior code.  We show how our dataflow analysis allows us to focus on relevant code at a point of completion, which when combined with local program context prior to the function call produces much better code completion performance than the context alone.
     
     \item[Automated Machine Learning] which takes a given dataset in the form of a structured table, and creates an effective machine learning pipeline to learn to predict some columns based on other columns.  Prior approaches have been based on dynamic analysis, and we show static analysis does just as well.  Static analysis is more practical, as actually running these pipelines is an arduous and expensive task; and one can mine large open repositories to populate such databases using analysis.
\end{description}
 
In the rest of the paper, we first describe Serenity's techniques for modeling Python based on a running example (Sections \ref{sec:running_example} and \ref{sec:analysis}). We then validate our analysis with two applications in code completion (Section \ref{sec:code_completion}) and automated machine learning (AutoML) (Section \ref{sec:automl}). We finally survey related work in Section \ref{sec:relatedwork} and conclude in Section \ref{sec:conclusion}. 
\section{A Running Example}
\label{sec:running_example}

{
\makeatletter\newcommand{\miniscule}{\@setfontsize\miniscule{5}{7}}%

\renewcommand\theFancyVerbLine{%
\ifnum\value{FancyVerbLine}=12
  \setcounter{FancyVerbLine}{15}{\miniscule \arabic{FancyVerbLine}}
\else\ifnum\value{FancyVerbLine}=27
  \setcounter{FancyVerbLine}{41}{\miniscule\arabic{FancyVerbLine}}
\else\ifnum\value{FancyVerbLine}=43
  \setcounter{FancyVerbLine}{71}{\miniscule\arabic{FancyVerbLine}}
\else{
  \miniscule \arabic{FancyVerbLine}}
\fi
\fi
\fi
}

\begin{figure}
\begin{minipage}{.8\columnwidth}
\begin{minted}[linenos, breaklines, fontsize=\small, ,escapeinside=!!]{python}
debug_level = 3!\label{re:debug_start}!

if debug_level > 5:
    def fd(model, test, train):
        assert len(test) < 500
        model.fit(train, test)
    fitit = fd
    
else:
    fitit = lambda model, test, train: model.fit(train, test)!\label{re:debug_end}!
    
from sklearn.svm import LinearSVC!\label{re:import_svc}!

from pystruct.models import MultiClassClf
from pystruct.learners import (NSlackSSVM, OneSlackSSVM, SubgradientSSVM, FrankWolfeSSVM)

digits = load_digits()!\label{re:load}!
X, y = digits.data, digits.target!\label{re:read_data_target}!
X = X / 16.!\label{re:manip_data}!
X_train, X_test, y_train, y_test = train_test_split(X, y)!\label{re:train_test}!

# we add a constant 1 feature for the bias
X_train_bias = np.hstack([X_train, np.ones((X_train.shape[0], 1))])!\label{re:x_train}!

fw_bc_svm = FrankWolfeSSVM(model, C=.1, max_iter=50)!\label{re:FrankWolfeSSVM}!

libsvm = LinearSVC(multi_class='crammer_singer', C=.1)!\label{re:make_svc}!
start = time()
fitit(libsvm, X_train, y_train)!\label{re:fit_svc}!
time_libsvm = time() - start
print("Score with sklearn and libsvm: %f (took %f seconds)" % (libsvm.score(X_test, y_test), time_libsvm))

start = time()
fitit(fw_bc_svm, X_train_bias, y_train)!\label{re:FrankWolfeSSVM_fit}!

\end{minted}
\end{minipage}
\caption{A Running example}
\label{running_example}
\end{figure}
}

Figure~\ref{running_example} shows the running example for this paper, a snippet of Python code adapted from the \texttt{multi\_class\_svm.py} script of ETH150K \cite{eth150k} benchmark.  Our modification (lines \ref{re:debug_start} to~\ref{re:debug_end}) is to add some debugging options to illustrate complexities in Python.  The variable \texttt{fitit} is assigned one of two functions depending on \texttt{debug\_level}: either a closure or a function with an added assertion.  The code loads a digits dataset (line \ref{re:load}), reads specific fields of the dataset (line \ref{re:read_data_target}), manipulates the dataset (line \ref{re:manip_data}), splits the data into train and test splits (line \ref{re:train_test}), and finally creates \texttt{X\_train\_bias} (line~\ref{re:x_train}.  The code then creates machine learning models \texttt{FrankWolfeSSVM} (line~\ref{re:FrankWolfeSSVM}) and \texttt{LinearSVC} (line~\ref{re:make_svc}).  The code then calls them: a \texttt{fitit} call on a \texttt{LinearSVC} (line~\ref{re:fit_svc}) takes the model (line \ref{re:x_train}), and \texttt{fitit} is called on \texttt{FrankWolfeSVC} with \texttt{X\_train\_bias} (line \ref{re:FrankWolfeSSVM_fit}).  Note that the call to the constructor \texttt{FrankWolfeSSVM} is on line \ref{re:FrankWolfeSSVM}, and the \texttt{fitit} call on the object is on line \ref{re:FrankWolfeSSVM_fit}, reflecting a property of most code - it is non-local.

\section{Analysis}
\label{sec:analysis}
\subsection{Background: call graph framework}
\label{sec:framework_intro}

Grove et al.~\cite{10.1145/506315.506316} provide a framework for expressing call graph algorithms for object-oriented languages.  It encapsulates the bulk of the algorithm, parameterizing the algorithms with functions that determine how to add context sensitivity.  The details of the framework are beyond the scope of this paper, but we depend on two details:
\begin{itemize}
\item First, we will rely later on something called the Procedure Key Selection Function (PKS), which is essentially a way to specify when called functions should be analyzed in a context-sensitive manner.  \item Second, the framework distinguishes between function call sites and object creation sites, which, as we shall see in Figure~\ref{fig:dynamic_code_examples}, is not possible in general in Python.  Hence, we combine the two sets into a single one.
\end{itemize}

The framework paper defines relevant program features at the top of page 694, which we excerpt here in Figure~\ref{fig:grove-defns}.  We need two minor changes:
\begin{description}
\item[InstVariables] is taken to be the set of strings possibly used as field names, rather than a set of declared field names, which it is in the original framework.  While field names can be defined in Python, this is entirely optional so we ignore such definitions.
\item[NewSites] becomes the same as the set {\em CallSites} to effect the second item above; that is, there is one set that combines all possible call sites and creation sites.  This represents the fact that every site can potentially see both classes and functions.
\end{description}

\begin{figure}[htb]
\begin{minipage}{.8\columnwidth}
\centering
\begin{description}
    \item[Class] all class declarations in the program
    \item[InstVariable] all instance variable declarations of the program
    \item[Procedure] all procedure declarations of the program
    \item[Variable] all variable names used in the program
    \item[CallSite] all call sites in the program
    \item[NewSite] all new sites in the program
    \item[LoadSite] all loads of instance variables in the program
    \item[StoreSite] all stores to instance variables in the program
    \caption{Program features from~\cite{10.1145/506315.506316}}
    \label{fig:grove-defns}
\end{description}
\end{minipage}
\end{figure}

\subsection{Language modeling}
\label{sec:language_model}

\begin{figure}[htb]
\begin{center}
\begin{minipage}{.8\columnwidth}
\begin{minted}{python}
import sys

if sys.argv[1] == "class1" or sys.argv[1] == "inst":
   class X:!\label{cls1}!
      pass
  
  if sys.argv[1] == "inst":
      X = X()

elif sys.argv[1] == "class2":
   class X:!\label{cls2}!
      def __new__(*args):
         return 0

elif sys.argv[1] == "def1":
   def X():!\label{def1}!
      return 1

elif sys.argv[1] == "def2":
   X = lambda: 2!\label{def2}!

elif sys.argv[1] == "import":
   import X!\label{import}!

elif sys.argv[1] == "method":

     if sys.argv[2] == "static":
        class X:!\label{classs}!
            def s():
                return 5

        X = X.s!\label{staticm}!
          
     elif sys.argv[2] == "instance":
        class X:!\label{classi}!
            v = 4
            def i(self):
                return self.v

        y = X()
        X = y.i!\label{instm}!
        
        
print(str(X))
print(str(X()!\label{dynamic_call}!))
\end{minted}
\end{minipage}
\end{center}
\caption{Dynamic code examples}
\label{fig:dynamic_code_examples}
\end{figure}

Figure~\ref{fig:dynamic_code_examples} illustrates the kind of dynamism with which analysis of Python must contend, in this case 5 different options for the meaning of \texttt{X()} on line~\ref{dynamic_call} based on the value supplied as \texttt{sys.argv[1]} and sometimes \texttt{sys.argv[2]}:
\begin{description}
 \item[class1] \texttt{class X} (line~\ref{cls1}) defines an ordinary class named \texttt{X}, of which line~\ref{dynamic_call} creates an instance.
 \item[class2] \texttt{class X} (line~\ref{cls2}) defines a class named \texttt{X} that redefines the \texttt{new} operator, so line~\ref{dynamic_call} just returns 0.
 \item[def1] \texttt{def X} (line~\ref{def1}) defines a function named \texttt{X}, and calling it at line~\ref{dynamic_call} returns 1.
 \item[def2] \texttt{X = lambda\dots} (line~\ref{def2}) creates a closure and assigns it to \texttt{X}; calling the closure at line~\ref{dynamic_call} returns 2.
 \item[import] The module \texttt{X} (line~\ref{import}) overrides default module behavior to become callable and return 3 at line~\ref{dynamic_call}.
 \item[method static] \texttt{X} (line~\ref{staticm}) is assigned the static method \texttt{s} of class \texttt{X} (line~\ref{classs}) which returns 5 at line~\ref{dynamic_call}.
 \item[method instance] \texttt{X} (line~\ref{instm}) gets a bound instance method (i.e. a closure over \texttt{y}) \texttt{i} of class \texttt{X} (list~\ref{classi}), returning 4 at line~\ref{dynamic_call}).
\end{description}
Note that all of these definitions of \texttt{X} can flow to the same call at line~\ref{dynamic_call}, so there is literally no syntactic distinction between different kinds of allocations, calls, and even modules in some cases.  And class and function names are all first class.  Thus analysis must handle these basic operations in a dynamic manner, unlike e.g. Java, where calls, allocations and imports have clear syntactic distinctions.  Note further that even basic method calls require closures to handle line~\ref{instm}.
 
The \texttt{X()} at line~\ref{dynamic_call} is a call on \texttt{X}, and this allows us to use standard dynamic dispatch to model all of this behavior, using synthetic "methods" where needed to handle language semantics.  We will use a similar "dispatch" at field accesses to handle the difference between class and instance fields, which again can only be known from the object accessed.  We shall make use of these indirections to define our framework model in Section~\ref{sec:framework_model}.

We adopt the terminology of Grove et al.~\cite{10.1145/506315.506316} to present our work as extensions to standard object-oriented call graph construction.  To fit our dynamic Python context, we make a few changes to the core definitions of that work.  
These changes reflect that Python does not require that fields be declared in order to be used, and it makes no syntactic distinction between calls and allocations.  Furthermore, as is standard for representing first-class entities in an object-oriented framework, we have one class for each first-class entity.  As Figure~\ref{fig:dynamic_code_examples} shows, classes, functions, methods and modules are all first-class, so our set of classes for analysis includes the following:
\begin{description}
    \item[$C_{class}$] a class representing program class \texttt{C}
    \item[$C_{inst}$] a class representing instances of class \texttt{C}
    \item[$M_{inst}$] a class representing instances of module \texttt{M}
    \item[$D_{inst}$] a class representing instances of function \texttt{D}
    \item[$S_{inst}$] a class representing the instance of script \texttt{S} 
\end{description}
Now most of the irregularities of Python calls and creations are handled by treating every call site as a {\em CallSite} for each receiver type $*_{inst}$ and as a {\em NewSite} for every receiver type $*_{class}$.  The site on line~\ref{dynamic_call} in Figure~\ref{fig:dynamic_code_examples} would have some types handled by each mechanism.  Fields are also handled seamlessly: on line~\ref{staticm}, \texttt{X} is a $C_{class}$, and on line~\ref{instm} \texttt{X} is a $C_{inst}$, so static and instance state are handled by making static fields be instance fields of the class object.

Call graph construction starts with a root stub that creates an instance of the main script $S_{inst}$ and calls it.  

\subsection{Framework modeling}
\label{sec:framework_model}

In many situations, it is difficult or impossible to find actual code for Python \texttt{import}s: there is no fixed relationship between names in \texttt{import} statements and locations of actually source code.  Even if there were, the structure of Python libraries is such that large amounts of the code is native and hence a Python analysis framework is not applicable.  Even if it were possible to find Python code, many libraries are large enough to make precise analysis challenging.  In our case, we are interested in the behavior of application code rather than library internals, so we minimize these issues by largely not analyzing framework code.

 Our model, called Turtles\footnote{from "turtles all the way down".  This phrase is of unknown origin, see \url{https://en.wikipedia.org/wiki/Turtles_all_the_way_down}}, abstracts Python frameworks to capture how the framework interacts with user code and to ignore all of its internal details.  Specifically, we model four aspects, all using the indirections of Section~\ref{sec:language_model}:
 \begin{enumerate}
    \item \label{model_import} We model import statements as returning a new framework, denoted by the name of the imported module. The framework is an opaque object with no functionality beyond implementing the model.
    \item \label{turtle_call} Calls to framework functions and methods typically return something, which is then possibly used by the user code.   We model every call to the framework as returning a new object from it; this model is transitive, so calls on those objects return further new objects from the framework.  We label these objects with the path by which they are accessed.
    \item \label{turtle_read} Accesses to fields of framework objects have little meaning in our model since we do not model the framework state at all.  However, user code typically expects that a field access return something, so we model all such field accesses as returning the container object.
    \item \label{turtle_arg} Arguments to turtle methods are mostly ignored, since we do not model what the framework does to them; however, sometimes functions are passed as parameters, and we assume that the framework might call it.  Since we do not model internal framework state, the model invokes callbacks from where they are passed as arguments.
 \end{enumerate}
 
The framework of Grove et al.~\cite{10.1145/506315.506316} provides the customization support needed to implement this model.  We start by introducing a new type of class, $T_{path}$, that represents a turtle, i.e. an opaque model object.  Item~\ref{model_import} is implemented by modeling \texttt{import M} statements as a call to a synthetic import procedure with \texttt{M} as its argument.  This call is modeled as returning a $T_{M}$.  Item~\ref{turtle_call} is implemented as a Procedure Key Selection Function (PKS) which takes the receiver of a type $T_{path}$ and the name $n$ of the called procedure and returns a new turtle of $T_{path.n}$.  Item~\ref{turtle_read} is implemented by simply returning self when reading any field of any $T_{path}$ type.  Item~\ref{turtle_arg} is implemented as a PKS that generates calls for every argument that is of a function type (this is not illustrated in our example).

\subsection{Inheritance from Turtles}

One wrinkle in our data is that application classes often inherit from turtle classes, meaning that method calls on \texttt{self} should logically be turtle methods when the method read is never assigned.  That is, if a read of \texttt{self} is to a field or method that is never assigned and the class inherits from a turtle, the read should return a new turtle object to capture unknown superclass behavior.  However, this is tricky to do because, since methods and fields can be assigned anywhere in the code, it is not in general possible to know if one will not be assigned until analysis terminates.  What we need to do is record such reads and, when analysis terminates, process them as turtle reads and restart analysis.  This restarting itself may need to be repeated, since reading one turtle could make more code reachable.

\subsection{Analysis of running example}

 \begin{figure}[hbt]
    \centering
    \includegraphics[width=.8\columnwidth]{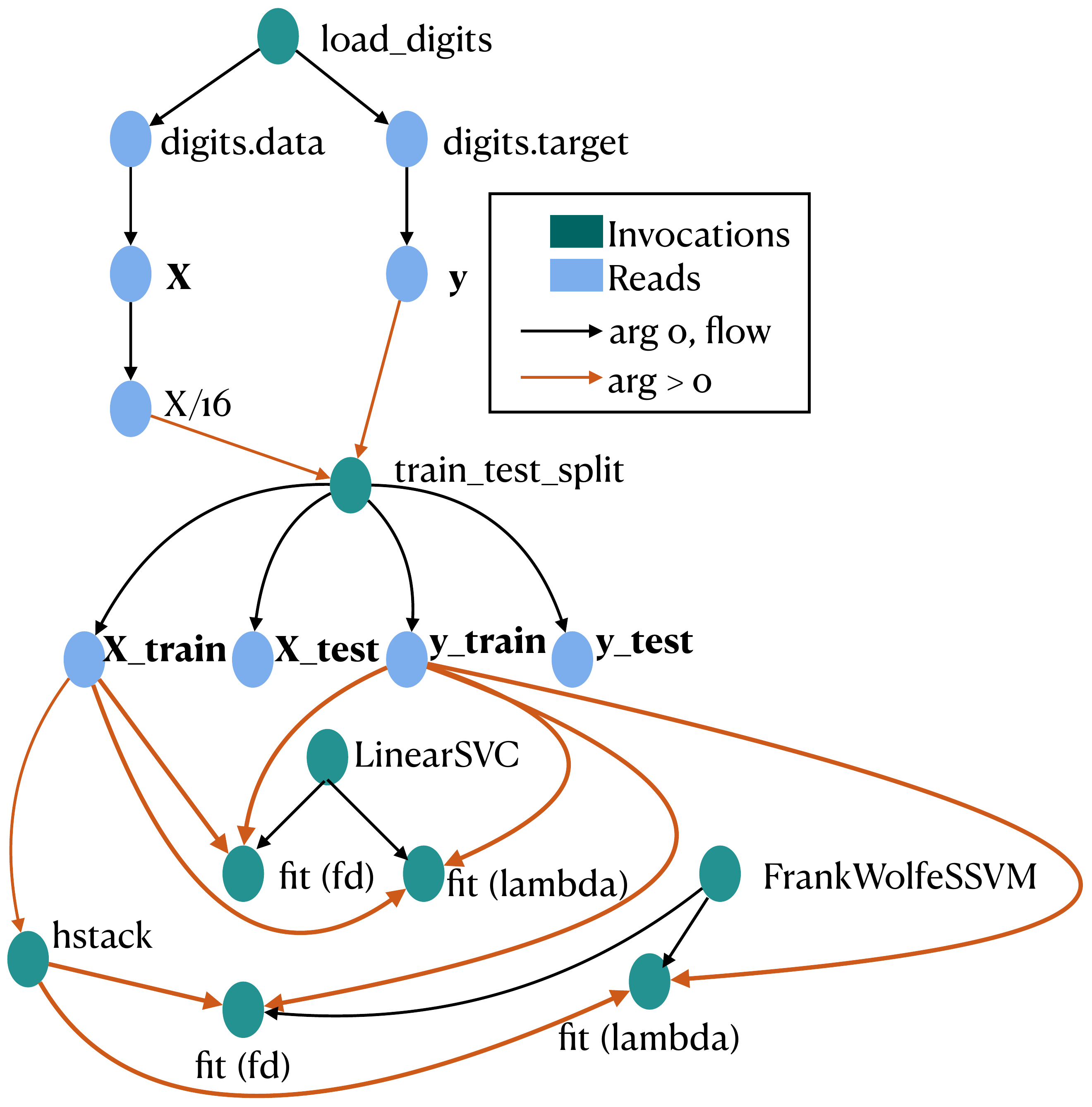}
    \caption{Dataflow graph for the running example}
    \label{fig:dataflow_graph}
\end{figure}

When this analysis is applied to the running example (Figure~\ref{running_example}), the result is the dataflow graph shown in Figure~\ref{fig:dataflow_graph}.  To illustrate our framework model, observe the import call of \texttt{LinearSVC} on line 15; as an \texttt{import}, this returns an object of type $LinearSVC_{inst}$, that is, an instance of the module.  When this is called (line~\ref{re:make_svc}), it returns a turtle of type $T_{LinearSVC}$, illustrated by the green node labeled \texttt{LinearSVC}.  When \texttt{fit} is called on this object in the \texttt{fitit} functions (line~\ref{re:fit_svc}), item~\ref{turtle_call} means it returns a derived turtle of type $T_{LinearSVC.fit}$, shown as a green node labeled \texttt{fit}.  Since \texttt{fit} is called on \texttt{LinearSVC}, a black data flow edge connects them.  On the other hand, the other non-\texttt{self} arguments to \texttt{fit} are shown with red arrows.  Other turtle functions are shown similarly: \texttt{load\_digits}, \texttt{train\_test\_split}, \texttt{hstack}, \texttt{FrankWolfeSSVM}.  Note that analysis has no idea what these functions do, just that they pass data.   Note that \texttt{fitit} is a variable holding one of two first-class functions, and it is called for both of the ML models created.  To get the precise results shown in Figure~\ref{fig:dataflow_graph} requires analysis infrastructure that handles first-class functions and also does context-sensitive analysis.  In particular, the model objects and the data flow to both the normal and debugging functions assigned to \texttt{fit}, since both potentially flow to \texttt{fitit}.  In the figure, the nodes are distinguished with labels of the function in which they occur.  

Other nodes in Figure~\ref{fig:dataflow_graph} represent local dataflow.  The topmost two blue nodes represent reads of the \texttt{data} and \texttt{target} fields of \texttt{data}, so they have edges from the \texttt{load\_digits} call and edges to their respective variables \texttt{X} and \texttt{y}.  \texttt{X} is scaled by \texttt{16}, shown by the nodes labeled \texttt{X/16}.  \texttt{X/16} and \texttt{y} then flow to \texttt{train\_test\_split} with red edges since they are arguments.

This graph focuses on data flow, which captures patterns of how the various turtle APIs are used across programs.  This allows us to learn patterns that enable our applications.

\subsection{Implementation}

Our analysis is implemented using WALA and its support for both Python 2 and Python 3 using the Jython system.  WALA is built to be extensible, and we used several features to ease our implementation work.  

The main extension is for handling turtles.  For item~\ref{model_import}, we override the model function that handles \texttt{import} to return a synthetic object with a turtle type named for the given module.  For item~\ref{turtle_call}, we override the selection of called methods for turtle classes so that any call goes to a synthetic method that creates and returns a turtle with the appropriate extended turtle name.  For item~\ref{turtle_call}, this synthetic method mostly ignores its arguments, except generating a call to each one to handle callbacks.  For item~\ref{turtle_read}, we override the code handling field reads to simply return the container if it is of turtle type.  

The other configuration is to add aggressive context  sensitivity for all turtle types.  Since the synthetic methods are trivial anyway, it is cheap to ensure that every call site is analyzed separately.

\section{Code Completion Application}
\label{sec:code_completion}
The core research question we ask is how useful Serenity's analysis is and whether it can help other applications, despite the challenges in modeling dynamic languages such as Python accurately.  
As a first application, we examine a code completion use case, which we cover below in detail. 
By code completion, we refer to the problem where, when given a snippet of a program, the problem is to predict a function call, analogous to what an IDE does for method suggestions.  We do not refer to code generation given natural language descriptions of code requirements, as in the Codex model that powers GitHub Co-Pilot \cite{chen2021codex} or even models that generate entire functions in a generative style based on function signatures or snippets of code, such as CodeT5 \cite{wang2021codet5}.  Our observation is that for code completion, the analysis requirement is that the methods be callable from a specific type, and so analysis for code completion is focused on detecting the types of objects.  For languages such as Python, type inference is hard, but our hypothesis is that code completion can benefit significantly from the data flow analysis that Serenity produces, simply because data flow can provide a focused context for code completion.

Recently, there have been a plethora of neural models of code such as \cite{codebert}, \cite{graphcodebert}, \cite{cubert}, \cite{wang2021codet5} trained with the objective of either predicting randomly masked tokens in code, or predicting the very next token, which one might assume is consistent with the task of code completion.  Our research question is whether one can leverage the extensive training of these models on millions of programs to perform code completion.  Specifically, we asked whether data flow analysis provided by Serenity can improve code completion when combined with these neural models.  If data flow analysis does provide any signal from Serenity, it should improve performance on code completion task even with the extensive training these language models already had.  We therefore modeled code completion as a fine tuning task, and varied the training inputs of fine tuning to be one of the three conditions shown below:
\begin{itemize}
\item All code as text prior to the function call
\item A slice of the code restricted to source expressions that are relevant to a function call in data flow 
\item Both code as text, as well as the slice, separated by a token to distinguish the two inputs.
\end{itemize}

\begin{figure}
\begin{minipage}{.8\columnwidth}
\begin{minted}[breaklines, linenos, fontsize=\small]{python}
print("Score with pystruct subgradient ssvm: %f (took %f seconds)" % (np.mean(y_pred == y_test), time_subgradient_svm))

# the standard one-vs-rest multi-class 
# would probably be as good and faster
# but solving a different model
libsvm = LinearSVC(multi_class='crammer_singer', C=.1)
start = time()
libsvm.fit(X_train, y_train)
time_libsvm = time() - start
print("Score with sklearn and libsvm: %f (took %f seconds)" % (libsvm.score(X_test, y_test), time_libsvm))


start = time()
fw_bc_svm.?
\end{minted}
\end{minipage}
\caption{Code snippet used for prediction}
\label{prediction_text}
\end{figure}

\begin{figure}
\begin{minipage}{.8\columnwidth}
\begin{minted}[fontsize=\small, linenos, breaklines]{python}
from sklearn.cross_validation import train_test_split
from pystruct.models import MultiClassClf
from pystruct.learners import (NSlackSSVM, OneSlackSSVM,
digits = load_digits()
digits.data
digits.target
X = X / 16.
train_test_split(X, y)
X_train_bias = np.hstack([X_train, np.ones((X_train.shape[0], 1))])
model = MultiClassClf(n_features=X_train_bias.shape[1], n_classes=10)
fw_bc_svm = FrankWolfeSSVM(model, C=.1, max_iter=50)
fw_bc_svm.?
\end{minted}
\end{minipage}
\caption{Code snippet corresponding to a slice from the analysis graph}
\label{training_slice}
\end{figure}
For all text code prior to a function call, there are limits on how many tokens modern language models can fit.  That is, when the code goes beyond the limit, truncation is needed in order for the models to run.  A widely-used truncation strategy is to only keep $n$ tokens prior to the prediction point, where $n$ is the maximum sequence length, which can lead to fairly local information, as shown in Figure~\ref{prediction_text} for our running example shown in Figure~\ref{running_example}.  The key prediction in Figure~\ref{prediction_text} is to predict what method will be called on \texttt{fw\_bc\_svm}, but notice that the construction of \texttt{fw\_bc\_svm} is out of the scope of the truncation\footnote{In this example, truncation was set to 1024 tokens, as per the requirements of one of the CuBERT models \cite{cubert}}.

For obtaining the slice restricted purely to dataflow, given a program and its corresponding dataflow graph, to predict the function call, we start at a node that we would like to predict, reverse all edges coming into the node, and find all reachable nodes.  Each node in the reachability set corresponds to a source expression in the original program, and we only include the expressions that are not sub-expressions of any other expressions as features.  Then, we order these expressions according to their positions in the source files, and add in variable names from the analysis artifacts so the code looks more or less like real code that the language models have been trained on.  Figure~\ref{training_slice} shows an example of such a dataflow based slice looks for the code in Figure~\ref{running_example}.  Here we start the \texttt{fw\_bc\_svm.fit} call in Figure~\ref{fig:dataflow_graph}, reverse all edges coming into the node, and perform a reachability analysis, to gather the slice, adding variable names such as \texttt{digits = load\_digits()}.  In this example, dataflow analysis does give important information useful for predicting the function call, because the slice brings in non-local but relevant code such as the definition of \texttt{fw\_bc\_svm} into the scope of text that can be fed to a neural model.


\subsection{Dataset} We used the popular benchmark of ETH150K \cite{eth150k}, which comes with 100K programs used for training, and the remaining 50K used as a testing set. ETH150K was analyzed using Serenity, and 147,288 of 150,000 files were successfully analyzed.  For the analyzed files, we parsed each file with a Python AST parser, and gathered all function calls.  For each function call identified by the AST, we examined whether we could find the function in the analysis output, and if it was found in the output, we checked if the source location of the call matched that in the AST.  Our observation has been that the Jython source mappings can be wrong sometimes, so we used both metrics to measure the completeness of the analysis.  The analysis found 58.77\% of function calls in the AST with matching source locations, and 67.36\% of function calls when the requirements to match source was relaxed.  Manual inspection on a few cases where source locations did not match indicated that the problem was indeed mapping being incorrect in Jython.  Further investigation revealed that many of the missing calls are instances of Python primitives that Serenity does not model and treats as no-ops, such as \texttt{repr} and \texttt{FutureWarning}.  A small fraction was found to be genuinely dead code, especially when Python files were integral parts of a larger application, as they often are in ETH150K.

\begin{figure}
    \centering
    \includegraphics[width=.8\columnwidth]{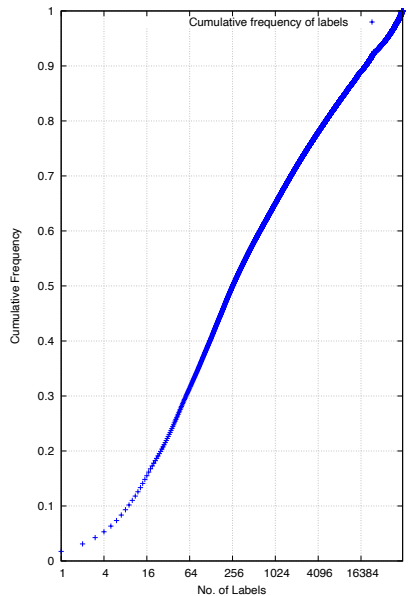}
    \caption{Distribution of labels for the classification task}
    \label{data_disbn}
\end{figure}

To generate the slices, we started with leaf nodes, and restricted ourselves to cases where the nodes had at least a depth of 1 when the edges were reversed.  We note that in a majority of cases, leaf nodes were actually expressions, as shown in the example code in Figure~\ref{expression_code}.  We ignored these in creating our dataset because we were focused on a problem that cannot be solved by a pure lexical analysis of code. When we restricted ourselves to nodes that were potentially function calls rather than expressions, we generated slices from 65.35\% of the programs where there existed at least one slice where the leaf node was likely a function call.  For the train and test sets of programs, we generated 334,415 slices and 162,847 slices respectively by iterating over all the leaf nodes in dataflow graphs.  Once we consider leaf nodes as nodes for our prediction, there were a total of over 65K labels that were generated across train and test sets for code completion.  Figure~\ref{data_disbn} plots the cumulative frequency distribution of labels against the number of labels.  As shown in the Figure, the distribution of labels follows the usual power law, but we note that the most popular label appeared across train and test only 1.7\% of the time, and the top 10 labels cumulatively appeared only 11.0\% of the time.  In other words, this is a difficult classification problem\footnote{We will make the datasets and code for all the work reported in this section available as open source.}.  We note that this method of declaring code completion is more realistic compared to other means for code completion (such as measuring next token prediction), in the sense that this is often the case that IDEs focus on.

\begin{figure}
\begin{minipage}{.8\columnwidth}
\begin{minted}[fontsize=\small, linenos, breaklines]{python}
    def f_Hp(self, pars, p, inpt, target):
        eps = 1E-6
        deriv = self.fprime(pars, inpt, target)
        offseted = self.fprime(pars + p * eps, inpt, target)
        return (offseted - deriv) / eps
\end{minted}
\end{minipage}
\caption{Example of code where a leaf node is an expression}
\label{expression_code}
\end{figure}

\subsection{Language model selection} 
To decide on the best neural model to use as a basis for our code completion experiments, we tested a number of code related language models including CodeBERT \cite{codebert}, GraphCodeBERT \cite{graphcodebert}, CuBERT \cite{cubert} and CodeT5 \cite{wang2021codet5}. 
CodeBERT\cite{codebert} is a bimodal model trained on datasets with natural language (NL) -programming language  (PL) pairs (e.g. documentation/code pairs) across six programming languages (Python, Java, JavaScript, PHP, Ruby, and Go). Similarly, GraphCodeBERT uses NL-PL pairs for pretraining a code language model, but based on local data flow graphs extracted from Abstract Syntax Trees.  CuBERT \cite{cubert} is another BERT-based model fine-tuned on multiple classification tasks such as checking the presence of certain bugs and predicting exception types. CuBERT is trained only on Python code, and furthermore uses language level tokens as inputs to the model.  CodeT5 \cite{wang2021codet5} is an encoder-decoder model based on T5 architecture \cite{t5}  with code-specific knowledge trained to distinguish which tokens are identifiers and recover them when they are masked out. CodeT5 is fine-tuned using multiple CodeXGLUE benchmarks including understanding tasks such as code defect detection and clone detection, and generation tasks like code summarization and translation.

\begin{figure}
    \centering
        \includegraphics[width=1\columnwidth]{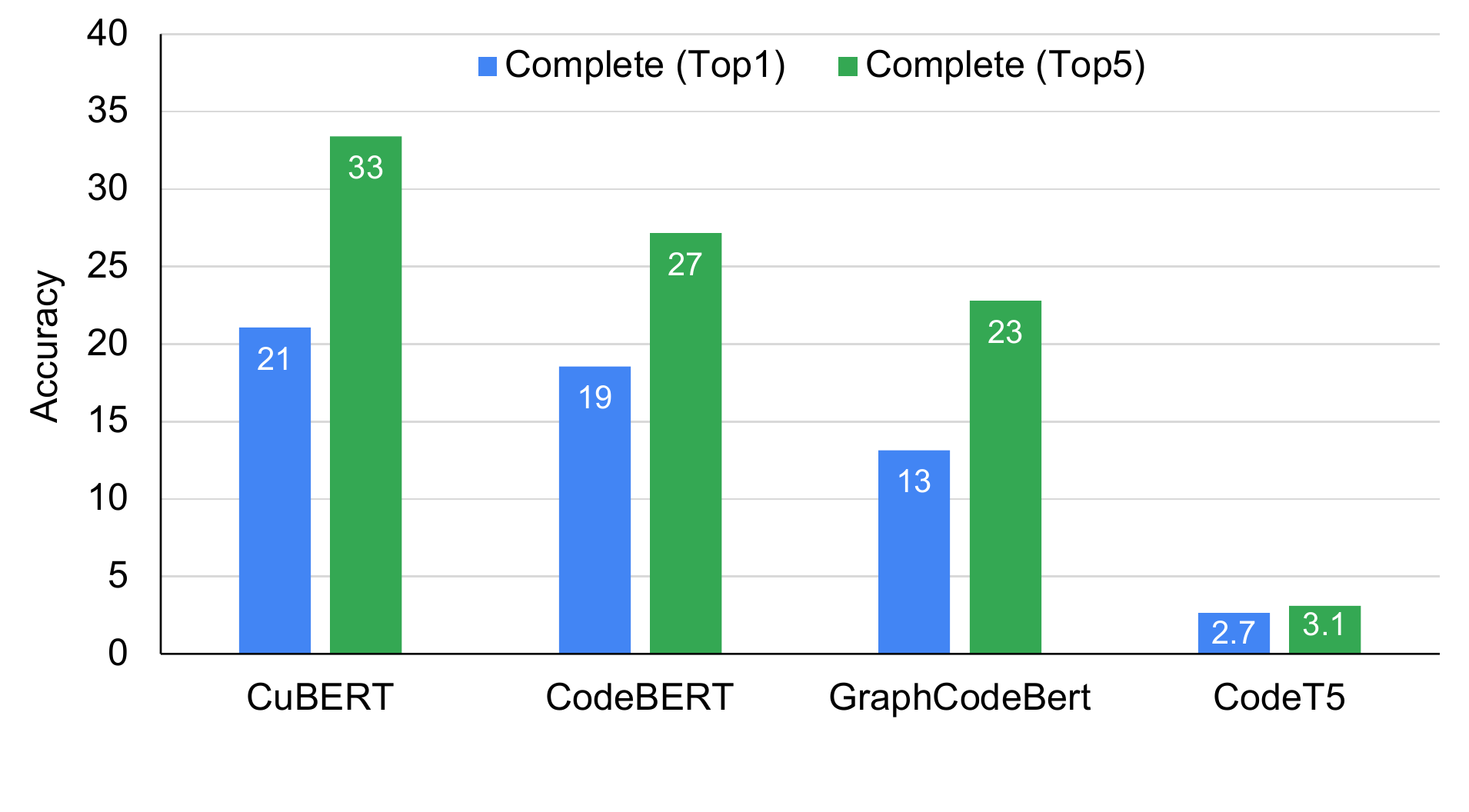}

    \caption{Accuracy on top-5 and top-1 test data for base language models.  Performance on CuBERT for slices is based on 150,739 and 137,987 examples for complete and slice, respectively because of tokenization issues.  For all other systems the number of testing examples was 167,816.}
    \label{lm_performance}
\end{figure}

Figure~\ref{lm_performance} shows the performance of these different models on the code completion task with no fine tuning for the top-1 and top-5 cases.  We modeled code completion as a mask prediction task, with the function call to be predicted being the masked token.  As shown in the Figure~\ref{lm_performance}, the best performing model was CuBERT on this task, which is not surprising because it was the only model trained exclusively on Python and used language level tokens unlike the other models.  We note that the performance of CodeT5 was surprisingly poor, but we think this may in part be due to the fact that it is trained on NL-PL pairs and it is strictly a generative model, for which we needed to specify a length of generation.  It also needed the most tuning in terms of specifying different search strategies for final token prediction, so it is possible that we did not choose the optimal search strategy for it.  For our purposes though, we chose CuBERT as a base, primarily because we expected to benefit most from fine tuning.  We point out that given our label distribution, for the language model to provide even 21\% performance on complete for top-1 and 33\% for top-5 is quite good. 

We turn now to the problem of fine tuning CuBERT with training inputs to see if analysis does in fact improve code completion as we defined it.  Note that CuBERT's pretraining was performed by feeding the model the logical lines of 5 million programs - so at the minimal, some fine tuning for the code context where the function call is to be predicted is  needed.  As stated earlier, we contrasted three different training conditions: 
\begin{itemize}
    \item \textbf{complete}: where we gave the model text starting from the call, backwards, as shown in Figure~\ref{prediction_text} 
    \item \textbf{slice}: where we used a backwards slice as shown in Figure~\ref{training_slice}
    \item \textbf{combined} where the text from \textbf{complete} and \textbf{slice} were concatenated as input to the model using a separator token.
\end{itemize}

\begin{figure}
\begin{minipage}{.8\columnwidth}
\begin{minted}[fontsize=\small, linenos, breaklines]{python}
response.json.return_value = dict(response, total_count=3, limit=0, offset=0)
        projects = self.redmine.project.all()
        self.assertEqual(projects.limit, 0)
        self.assertEqual(projects.offset, 0)
        self.assertEqual(projects[0].id, 1)
        self.assertEqual(projects[1].id, 2)
        self.assertEqual(projects[2].id, 3)

    def test_offset_limit(self):
        response_with_limit_offset = {'total_count': 2, 'limit': 3, 'offset': 1, 'projects': response['projects'][1:3]}
        self.response.json.return_value = response_with_limit_offset
        projects = self.redmine.project.all()[1:3]
        self.assertEqual(projects.limit, 3)
        self.assertEqual(projects.offset, 1)
        self.assertEqual(projects[0].id, 2)
        self.assertEqual(projects[1].id, 3)

    def test_offset_limit_mimic(self):
        projects = self.redmine.project.all()[1:3]
        self.assertEqual(projects.limit, 3)
        self.assertEqual(projects.offset, 1)
        self.assertEqual(projects[0].?
\end{minted}
\end{minipage}
\caption{Code snippet where local text can help prediction}
\label{complete_better_than_df1}
\end{figure}
\begin{figure}
\begin{minipage}{.8\columnwidth}
\begin{minted}[fontsize=\small, linenos, breaklines]{python}
from tests import unittest, mock, Redmine, URL
Redmine(self.url)
projects = self.redmine.project.all()[1:3]
        self.assertEqual(projects[0].?
\end{minted}
\end{minipage}
\caption{Code where data flow lacks sufficient context}
\label{complete_better_than_df2}
\end{figure}

The test was on \textbf{complete} text, or \textbf{combined}.  We chose these conditions because we observed from examples that for the problem of code generation, data flow is not sufficient by itself.  Figure~\ref{complete_better_than_df1} shows such an example.  In this code, lines 5 and 15 contain the clue needed to make the prediction of \texttt{id}, but they are unrelated to the receiver for which the call is being made on line 22.  Yet, the local pattern of code has the same variable names, and the same set of calls are repeated across functions, suggesting that \texttt{id} may be a good candidate label.  By contrast, the corresponding slice contains minimal information as shown in Figure~\ref{complete_better_than_df2}, since the receiver \texttt{projects[0]} was defined just within the function \texttt{test\_offset\_limit\_mimic}. 

\begin{figure}
\begin{minipage}{.8\columnwidth}
\begin{minted}[fontsize=\small, linenos, breaklines]{python}
import sys
import logging
from functools import partial
from datetime import datetime
from abc import ABCMeta, abstractmethod
import json
from _config import AttrDict

__all__ = ['multikey_getter_gen', 'unescape_json', 'LogParser', 'JSONParser', 'LogLine',
           'AccessLog', 'CommonLogFormat', 'uWSGIParser']

def multikey_getter_gen(parser, keys, is_indices=False, delimiter="\t"):
    """Generator meta-function to return a function
    parsing a logline and returning multiple keys (tab-delimited)"""
    if is_indices:
        keys = map(int, keys)

    def multikey_getter(line, parser, keyset):
        data = parser(line.strip())
        return delimiter.join((unicode(data[k]) for k in keyset))

    def multiindex_getter(line, parser, keyset):
        data = parser(line.strip())
        return delimiter.join((unicode( data.by_index( idx-1, raw=True)) for idx in keys))

    if is_indices is True:
        # Field indices
        return ?
\end{minted}
\end{minipage}
\caption{Example of where complete text may have text relevant to the prediction, but  distant from call site}
\label{complete_inadequate}
\end{figure}

We note however that sometimes the slice can help even when the truncation does not cut off key information for prediction. Figure~\ref{complete_inadequate} shows one such example.  The predicted function is \texttt{partial} is imported in line 3, but the actual call is on line 28.  On the other hand, in Figure~\ref{df_focused}, the import is the only call prior to the line, so the slice can make relevant information proximal, such that the neural model can pay greater attention to proximal elements of the code.

\begin{figure}
\begin{minipage}{.8\columnwidth}
\begin{minted}[fontsize=\small, linenos, breaklines]{python}
partial = #!/usr/bin/env python #
from functools import partial
keys = map(int, keys)
        return ?
\end{minted}
\end{minipage}
\caption{Example of where dataflow is very focused}
\label{df_focused}
\end{figure}

\subsection{Model details} We use the CuBERT model released by~\cite{cubert}\footnote{The CuBERT model can be accessed at github.com/google-research/google-research/tree/master/cubert}, which has 24 layers with 16 attention heads and 1024 hidden units and was pretrained on 4M unique Python files on Github.
At fine-tuning, we set the batch size to 10 and trained the model using 8 Tesla V100 with 32GB memory.
The learning rate is 5e-5, and we gradually warmed up the learning rate for the first 300 gradient updates, which are the default values provided by the HuggingFace library~\cite{wolf-etal-2020-transformers}.  The training stops after 20 epochs, or ends after the evaluation accuracy hasn't improved for three epochs.  For the \textbf{complete} and \textbf{slice} models we used the 512 tokens model, and when we used \textbf{combined}, we used the 1024 tokens model such that the exact same tokens present in \textbf{complete} and \textbf{slice} could be used together along with the separator.

We apply CuBERT's tokenization to Python programs in ETH150K where the Python programs are first tokenized using the standard Python tokenizer (the \texttt{tokenize} module)\footnote{github.com/python/cpython/blob/main/Lib/tokenize.py}$^,$\footnote{We note that \texttt{tokenize} only outputs tokens for the code snippet that is free of any syntax errors; otherwise, it returns either IndentationError or TokenError. To predict function calls we often feed code that is incomplete, thus syntactically incorrect; therefore, we had to modify the original module so that it always returns what has been already tokenized thus far.}.  Then we further break down the program tokens into 49,558 subwords using subword tokenization~\cite{vaswani2017attention}, as performed by the cuBERT tokenizer.

\subsection{Results of fine tuning}
\begin{figure}
    \centering
    \includegraphics[width=1\columnwidth]{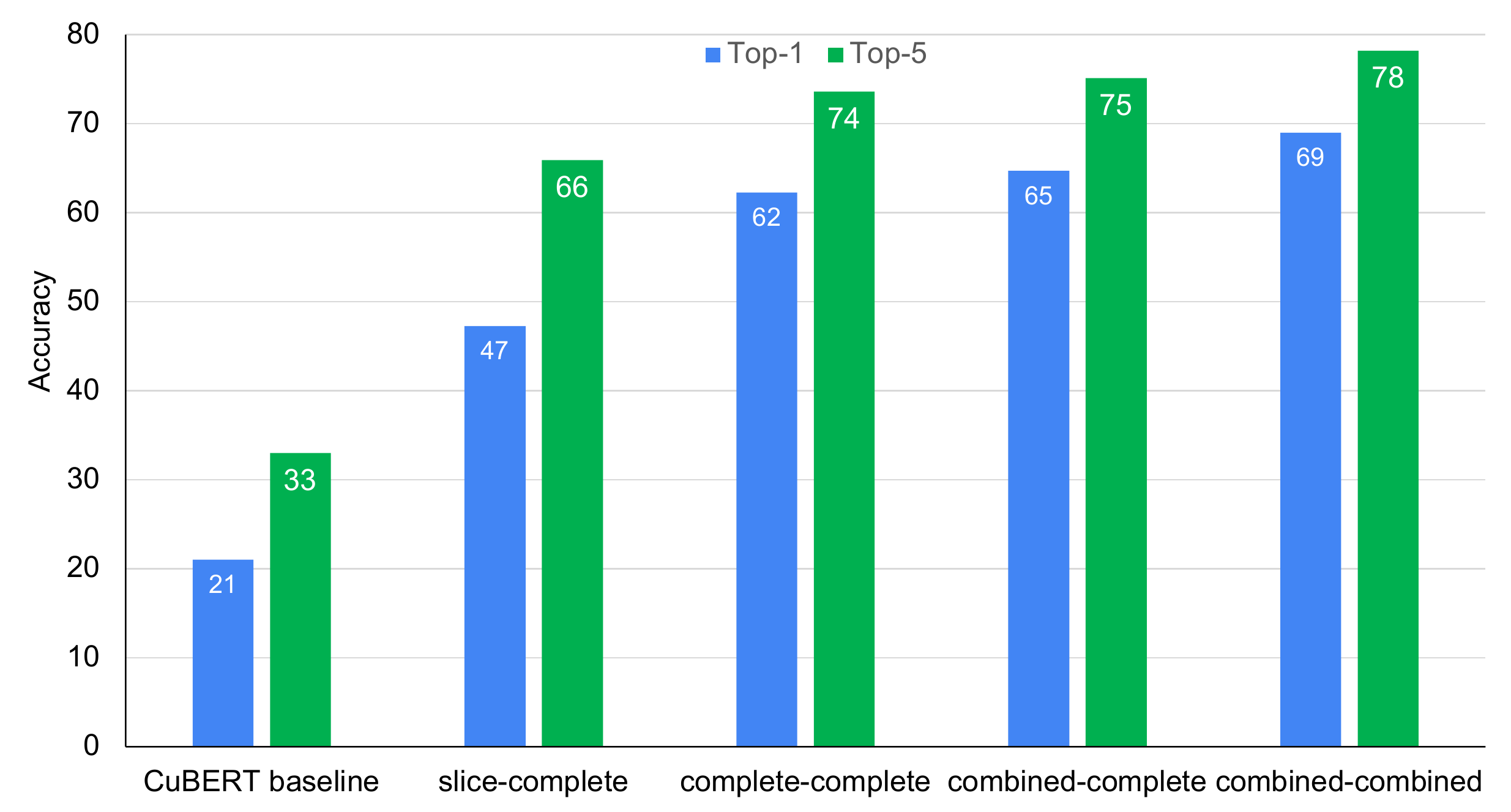}
    \caption{Results of fine tuning on the different "training - testing" conditions; i.e., training on \{slice, complete or combined\} and tested on \{complete or combined\}}
    \label{fine_tuning_results}
\end{figure}

Figure~\ref{fine_tuning_results} shows the accuracy in predicting the function call exactly across the different training and test conditions.  As shown in the Figure~\ref{fine_tuning_results}, training on \textbf{slice} was at 47\% accuracy when tested on the complete text (\textbf{slice-complete}), which is significantly above the 21\% of top-1 baseline from cuBERT.  Training on \textbf{complete} however was much better at 62\% on the same text (\textbf{complete-complete}), which is not surprising given that inspection of examples (e.g., Figure~\ref{complete_better_than_df1}) show that complete often contains the expressions in slice when the dataflow is local, and furthermore, benefits from repetition in coding patterns that might hint at labels in the absence of any real connection. The key question is whether slices provide any benefit over and above what benefit is gained from complete.  Training on combined suggests that slices do provide a strong signal, with a 65\% accuracy on complete text (\textbf{combined-complete}), and 69\% accuracy on the combined text (\textbf{combined-combined}).  We also compared top-5 performance across conditions to allow comparison to the baseline language models - not surprisingly this result improved accuracy across all conditions, with the \textbf{combined-combined} condition showing the best performance at 78\%.  The results show that data flow analysis can significantly augment code completion performance.

\begin{figure}
    \centering
    \includegraphics[width=.9\columnwidth]{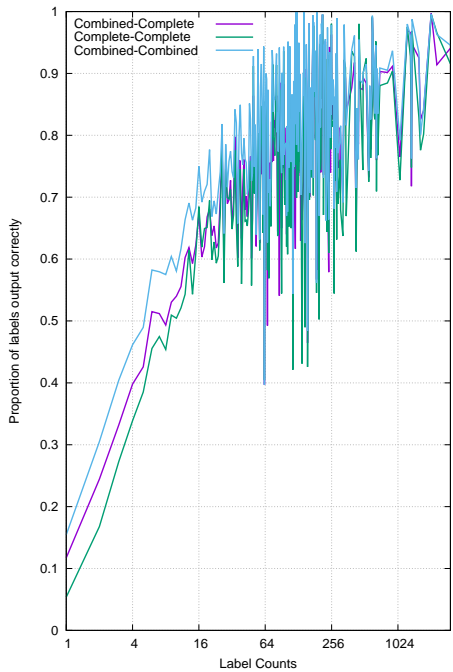}
    \caption{Accuracy on labels with different counts}
    \label{model_label_prediction}
\end{figure}

\subsection{How do slices help code completion?}
We conducted an analysis of how slices might help code completion performance; i.e., to understand if slices help the model complete code better for rare labels compared to more common ones, since statistical approaches likely work better for common labels, but less well for rare ones.  Figure~\ref{model_label_prediction} shows the performance of the different models; the presence of slices at training and test enhance code completion performance for rare labels more than common labels, although the advantage does seem to be present for common ones as well.  The \textbf{combined-combined} model was 15\% accurate on labels with count 1, of about 18,000 labels, and that number rapidly approaches 40\% for labels with count 3.  As labels become more frequent, the differences between the models gets more noisy but the \textbf{combined-combined} case still holds an advantage.

\subsection{Comparison to existing work}

In this space, comparisons are tricky because there is no common or standard benchmark and because the exact problem varies.  We chose ETH150K, which is at least a well-known code repository, but work that e.g. relies on its own sample of GitHub makes results incomparable.  The exact problem varies too, with some tools, like us, predicting function calls, others predicting only method calls and still others predict the next token for all tokens.  There is a real need for a benchmark in this space; as part of helping build such a benchmark, we will release our own slice and complete dataset to the community.
\cite{li2017code} reports overall accuracy numbers around 0.7, which is almost the same as ours, but that paper is predicting the next token across all token types, so could benefit from the fact that some predictions (e.g. ')' followed by ':' in \texttt{def}) follow from the grammar.  Pythia \cite{pythia} uses a neural network rather than a language model, but reports comparable accuracy numbers for top-1.  Their top-5 number is higher, but their predictions seem to be for method calls, rather than all functions as we do, for which the receiver may provide context to aid prediction.

These approaches may be complementary, too.  Our work showed that adding slice data to local context greatly aided the accuracy of our models.  Other approaches also rely on mostly local information, and could potentially benefit from slices as well.  We plan to investigate this further in our future work.


\section{Automated ML Pipelines Application}
\label{sec:automl}

\begin{figure}
    \centering
    \includegraphics[width=0.75\columnwidth]{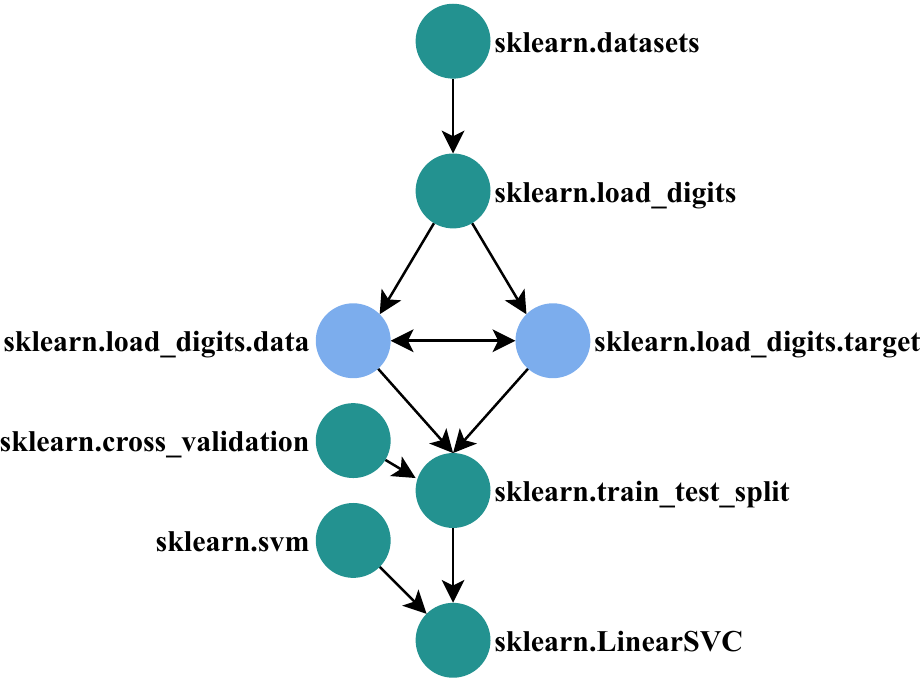}
    \caption{KGpip's training graph for our running example after filtering out  as input to the AutoML system.}
    \label{filtered_graph}
\end{figure}

The problem of automated machine learning pipelines (AutoML) focuses on automatically building pipelines by performing a search over valid data transformations and learners, along with hyper-parameter optimization for each learner. Our research question is whether we can perform learner and transformation selection based on mining large repositories of abstracted ML python scripts obtained statically by Serenity. Unlike dynamic analysis,  Serenity's static analysis of ML pipelines has the advantage of scaling to millions of scripts due to its low cost. Specifically, our question is whether the extracted semantics of a set of pipelines by Serenity can help in predicting a new pipeline when combined with neural graph models. 

\begin{figure*}
    \centering
    \includegraphics[width=1.5\columnwidth]{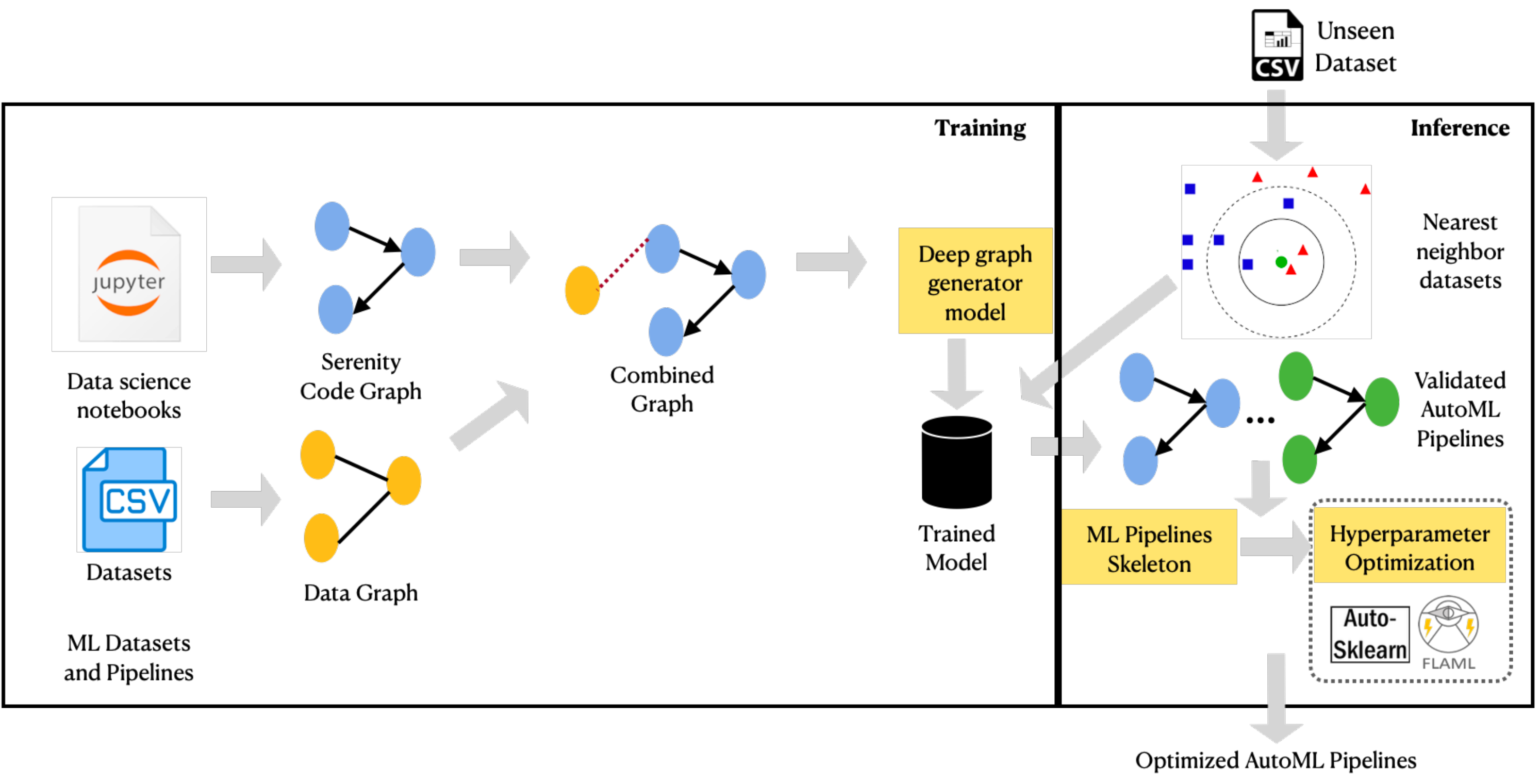}
    \caption{An overview of KGpip training and inference workflows}
    \label{kgpip_overview}
\end{figure*}

We developed an AutoML system, called KGpip, described in detail in a separate work ~\cite{kgpip}, which builds a database of datasets and their corresponding historically used ML pipelines using Serenity analysis. KGpip formulates the automation of a ML pipeline as a graph generation problem. It is based on two hypotheses that a neural graph generator will: i) capture more succinctly multiple pipelines seen in practice for a given dataset, and ii) capture statistical similarities between different pipelines more effectively. In KGpip,  we filter out Serenity's analysis to remove non-ML related components such as calls to libraries other than target ML libraries (Sklearn, XGBoost, and LGBM), and nodes indicating location of calls within a pipeline script, among others.  Figure~\ref{filtered_graph} shows the filtered version of the graph for our running example.  We also show in Figure \ref{kgpip_overview} an overview of how KGpip works at training and inference phases. Using code analysis graphs obtained from Serenity and dataset embeddings, KGpip trains a graph generation model optimized to output a ML pipeline as close as possible to the target pipelines of the training data. At inference time, KGpip identifies the closest dataset to the input dataset and uses its embedding as input to the graph generation model which in turn outputs a set of possible ML pipelines. These pipelines are then validated and fed to a hyper-parameter optimizer to get the best pipeline that results in the highest performance on the input dataset.

KGpip is designed to work with AutoML systems, such as  AutoSklearn \cite{autosklearn} and FLAML \cite{flaml}, to utilize their hyperparameter optimizers. With a collection of 2000 ML python scripts, we trained a graph generation neural network that learns to generate a ML pipeline graph for a given dataset. We conducted a comprehensive evaluation using 77 datasets from different benchmarks, such as AutoML and Penn Machine Learning Benchmark (PMLB), and different ML portals, such as Kaggle and OpenML. Table \ref{kgpip_performance} shows the overall KGpip performance which  significantly improves the selection of data transformation and learning algorithms of state-of-the-art AutoML systems, namely, Auto-Sklearn and FLAML. We note that AutoSklearn consults a database of pipelines and datasets, and picks pipelines to start the search based on a nearest neighbors to an unseen dataset, except that AutoSklearn's dataset consists of effective pipelines based on actual execution.  We also compared KGpip to AL~\cite{al}, which uses dynamic code analysis on existing machine learning pipelines to select optimized pipelines.  AL was unable to process 60 datasets because it ran out of time in searching for pipelines.  On a smaller set of 17/77 datasets on which AL was able to work, AL achieved an average performance of 0.36 compared to 0.745, 0.705, 0.79 and 0.765 by FLAML, Auto-Sklearn, KGpip + FLAML, and KGpip + AutoSklearn, respectively. This comparison with both AL and AutoSklearn clearly illustrates the value provided by Serenity, compared even to approaches that rely on dynamic runtime analysis. 


\begin{table}[]
\begin{tabular}{lll}
\toprule
                    & Average Performance & T-Test \\
\midrule
AutoSklearn         & 0.71 \footnotesize{(0.24)}        & -               \\
KGpip + AutoSklearn & 0.77 \footnotesize{(0.22)}        & 0.0002          \\
FLAML               & 0.71 \footnotesize{(0.27)}        & -                \\
KGpip + FLAML       & 0.77 \footnotesize{(0.20)}        & 0.0132         \\
\bottomrule
\end{tabular}
\caption{Performance (average and stdev) of KGpip compared to FLAML and AutoSklearn. Both variations of KGpip show significant improvements compared to existing systems, both with 2-tailed T-test $p < 0.05$ }
\label{kgpip_performance}
\end{table}

\section{Related Work}
\label{sec:relatedwork}

\noindent \textbf{Static Analysis for Python: }  
Static analysis of Python has attracted considerable interest lately, and there have been a range of approaches.  Type inference has been a focus of much work, some using techniques such as abstract interpretation and, more recently, there has been work using machine learning.  Likely the best-known work is MyPy~\cite{MyPy} and Pytype~\cite{Pytype}.  MyPy focuses on checking and inferring types that conform to PEP 484~\cite{vanrossum_lehtosalo_langa_2014}, which defines a syntax for Python types.  MyPy focuses on inference within a single function, since types are expressed at function boundaries in PEP 484.  Pytype does type inference, and it can handle cases where a variable has different types at different points.  It also does relatively little interprocedural analysis.  

Moat et al.~\cite{mine-ECOOP20} present an abstract interpretation for type inference of Python that models a variety of domains to compute more accurate information, and it makes use of the recency abstraction for aliasing.  However, it is currently limited in its support for interprocedural analysis, which is enabled by inlining.  Fritz and Hage~\cite{10.1145/3018882.3018888} present a dataflow analysis for type inference that provides a range of tradeoffs for cost and precision. It does handle features like first-class functions.

Machine learning is also used for type inference of Python. TypeWriter~\cite{typewriter} trained a neural model using a corpus of code, with labeled data derived from user annotations.  TypeWriter  considers comments in code as inputs to the neural model, unlike program analysis based type inference.  While such systems are certainly performing analysis, their approach and mechanism are quite different from ours.

There have been other analyses of Python, often for special purposes.  Ariadne~\cite{10.1145/3211346.3211349} makes use of WALA, as we do, but focuses on inferring the shapes of tensors in machine learning programs.  Unlike our approach to libraries, Ariadne models ML libraries as needed for tensor-related operations.

\noindent \textbf{Code completion}: Code completion has been a prominent area of research towards achieving better productivity when working within an IDE.
One of the main challenges in this domain is about code representation where the vast majority of work has used either tokens or abstract syntax trees to represent  code (we refer the reader to \cite{allamanis2018survey} for a detailed survey of this area).  \cite{li2017code} represents Python and JavaScript codes as ASTs and use a pointer network for better predicting Out of Vocabulary words in code completion. Pythia \cite{pythia} is another approach that uses ASTs with an LSTM model for code completion. 

A number of approaches also tried to leverage representations based on data and control flow \cite{DBLP:conf/iclr/AllamanisBK18,DBLP:journals/corr/abs-1811-01824, Chae:2017:AGF:3152284.3133925}. On JavaScript, \cite{DBLP:conf/icpp/HsiaoCN14} utilized a program dependence graph to detect code duplication. 
\cite{DBLP:conf/iclr/AllamanisBK18} use AST based representation augmented with local data and control flows for predicting variable names and variables misuse.  \cite{DBLP:journals/corr/abs-1811-01824} combines token based representations of code with edges based on object uses, and AST nodes to predict the documentation of a method. 
To perform code completion over Java API calls, \cite{Nguyen:2015:GSL:2818754.2818858, Nguyen:2009:GMM:1595696.1595767} used a mostly intraprocedural analysis for mining graphs augmented with control and data flow.  

With the rise of pre-trained language models such as BERT \cite{devlin2018bert} and GPT \cite{gpt2, gpt3}, many recent approaches \cite{codebert,cubert,graphcodebert,wang2021codet5,lu2021codexglue} started to leverage the already existing rich language understanding in these models and fine tune it for various code understanding tasks such as code summarization, translation, completion, bug detection, etc. CodeBERT\cite{codebert} is a BERT based model trained on pairs of natural language and programming language samples across six programming languages. GraphCodeBERT \cite{graphcodebert} uses BERT as well, but represents the code using data flow graphs based on ASTs.   The dataflow in GraphCodeBERT is completely local and not interprocedural, as in Serenity.  For instance as an example, it adds edges from all variables used in an expression to their definitions. CuBERT \cite{cubert} and CodeT5 \cite{wang2021codet5} are another two models based on BERT and T5 \cite{t5} architectures, respectively. 

Unlike our approach, all these methods represented code either as a sequence of tokens \cite{codebert,cubert}, ASTs \cite{maddison2014structured,wang2021codet5,pythia,li2017code}, or data flows derived from ASTs \cite{graphcodebert}. 



\noindent \textbf{AutoML approaches}: Several AutoML frameworks have been proposed recently \cite{flaml, autosklearn, al, drori2019automl}.  
In most AutoML systems, learner and pre-processing selection is driven by a database of actual executions of pipelines and data. For instance, \cite{autosklearn,Reif2012} compute a database of dataset meta-features such as number of rows and columns, while \cite{al} mines a repository of run-time information of inputs to the learners and preprocessors via dynamic code analysis of public ML pipelines available e.g. on Kaggle. The predicted learners and preprocessors are based on a similarity measure between the target dataset and stored features. In KGpip we utilized dense vector embeddings derived from raw contents of datasets to measure this similarity and graph neural networks to select the learners/preprocessors and generate the pipeline.

Some existing systems such as TPOT \cite{le2020scaling} or Recipe \cite{S2017RECIPEAG} use evolutionary algorithms for pipeline generation. Others approach it as a probabilistic matrix factorization \cite{NEURIPS2018_b59a51a3}, an AI planning problem when combined with a user specified grammar \cite{ICAPS20paper208,Ml-plan}, or a bayesian optimization problem combined with Monte Carlo Tree Search \cite{ijcai2019-457}. None of these approaches however use analysis to build up their database.

\section{Conclusion}
\label{sec:conclusion}
In this paper, we introduced Serenity; a framework for Python code static analysis. Serenity relies on two mechanisms (a) dynamic dispatching at the core of language translation, and (b) extreme abstraction of libraries. To demonstrate the utility of Serenity's analysis, we used it in two important code-related applications: code generation and automated machine learning. Serenity's analysis showed very promising performance in both applications, allowing us in some cases to outperform approaches based on dynamic analysis, and perform competitively for code completion. We also implemented Serenity as an open-source implementation based on WALA, a popular framework for program analysis. 


\bibliography{references}



\end{document}